\documentstyle[11pt,paspconf,epsf]{article}

\begin{document}
\title{Intracluster Medium Abundances Revisited}
\author{Brad K. Gibson \& Emma J. Woolaston}
\affil{Mount Stromlo \& Siding Spring Observatories, Australia} 

\begin{abstract}
We examine the origin of heavy elements in the intracluster medium (ICM) of
galaxy clusters, concentrating upon the roles played by 
supernovae (SNe) Types Ia and II.  The most accurately determined elemental
abundances, Si and Fe, imply a mild predominance of Type II SNe as the
source of ICM Fe, contributing $\sim 60\rightarrow 80$\% of its total (and
$\sim 100$\% of $\alpha$-elements).  (Currently) intractable uncertainties in
measuring X-ray $\alpha$-element ICM abundances, the 
initial mass function (IMF), and stellar evolution ingredients, make
a more precise determination of the Ia:II ICM iron ``ratio'' impossible.
\end{abstract}

\keywords{nuclear reactions, nucleosynthesis, abundances -- 
stars: supernovae -- intergalactic medium}

\section{Introduction}
Clusters of galaxies are embedded in a hot, iron-enriched,
X-ray emitting plasma.
While it is now generally accepted that this iron was
synthesized in $\sim$L$_\ast$
cluster ellipticals and ejected to the ICM via
SN-driven winds (Gibson \& Matteucci 1997; Renzini 1997), an unambiguous
\it a priori \rm prediction as to the primary 
synthesis source - i.e., Type Ia or Type II SNe - has proven difficult.  
The ambiguity arises due to the fact that both SNe types 
synthesize and eject iron to the interstellar medium (ISM), albeit on very
different timescales.
Models favoring an early-time Type II-dominated wind \it
and \rm those favoring a later-time Type Ia-dominated wind are both physically
plausible.

In principle, because the IMF-weighted Type II SNe yield ratios (i.e.,
[$\alpha$/Fe]) are
super-solar, whereas those of Type Ia SNe are $\sim 10\rightarrow 100\times$
underabundant with respect to the solar ratios, discriminating between these
two extreme models may be approached through determination of the ICM
[$\alpha$/Fe].  Unfortunately, as stressed by Mushotzky et al. (1996), the
spectral resolution afforded by present-day X-ray satellites makes this a
tricky proposition.  Despite the inherent difficulties, based upon 
high-quality \it ASCA \rm SIS spectra of four clusters, Mushotzky et al.
derived mean ICM abundance ratios of [Si,O,Mg/Fe]=+0.14,+0.01,-0.10 (as
scaled to the meteoritic abundances by Ishimaru \& Arimoto 1997).

In what follows, we pursue the implications of the Mushotzky et al. (1996) ICM
abundance ratio determinations, following the reasoning outlined in our earlier
study (Gibson, Loewenstein \& Mushotzky 1997; hereafter, GLM97).  We re-visit
the question of what these ratios tell us about the relative fractionary
contributions of Type Ia- versus Type II-iron to the total ICM iron.  
Attention is drawn to several
unappreciated points in both our earlier work (GLM97),
and that of Tsujimoto et al.
(1997; hereafter T97) and Thomas et al. (1998; hereafter TGB98).

\section{Stellar Yields}

First, let us echo the sentiments of GLM97 and TGB98 and state that not all
Type II SNe stellar yield compilations are created equal - see Figure 1 of
Gibson 1998a,b.  This can be further appreciated
through Table 1 below, in which we show the IMF-weighted elemental [O,Mg/Fe] 
(columns 2 and 4) and
isotopic [$^{16}$O,$^{24}$Mg/$^{56}$Fe] (columns 3 and 5) ratios, 
for both the Tsujimoto et al. (1995;
hereafter T95) and Woosley \& Weaver (1995; hereafter WW95) yields.

\begin{table}
\caption{IMF-weighted elemental and isotopic
abundance ratios for the T95 and WW95 Type II SNe stellar
yields (the superscript in Column 1 represents the grid metallicity [Fe/H]).  
An IMF of slope $x=1.35$, and Type II SNe-precursor range
$11\rightarrow 40$ M$_\odot$ was assumed.}
\begin{center}
\begin{tabular}{lrrrr}
\tableline
Grid        & [O/Fe] & [$^{16}$O/$^{56}$Fe] & [Mg/Fe] & [$^{24}$Mg/$^{56}$Fe] \\
\tableline
T95$^{+0}$       & +0.23$\;$ & +0.22$\quad$ & +0.25$\;\;$ & +0.22$\quad$ \\
WW95$^{+0}$      & +0.17$\;$ & +0.16$\quad$ & +0.08$\;\;$ & +0.04$\quad$ \\
WW95$^{-1}$      & +0.09$\;$ & +0.06$\quad$ & -0.07$\;\;$ & +0.00$\quad$ \\
WW95$^{-2}$      & +0.10$\;$ & +0.07$\quad$ & -0.12$\;\;$ & -0.04$\quad$ \\
WW95$^{-4}$      & +0.25$\;$ & +0.23$\quad$ & +0.08$\;\;$ & +0.17$\quad$ \\
WW95$^{-\infty}$ & -0.06$\;$ & -0.09$\quad$ & -0.19$\;\;$ & -0.12$\quad$ \\
\tableline
\tableline
\end{tabular}
\end{center}
\end{table}

\noindent
Columns 2 and 4 reiterate the points made by GLM97, TGB98, and
Gibson (1998a,b) that the T95 IMF-weighted yield ratios are significantly
higher (i.e., $\sim 15\rightarrow 50$\% greater)
than those found employing the WW95$^{+0}$ yields.  The metallicity-dependent
WW95 yields show a factor of two range in their IMF-weighted ratios.
Columns 3 and 5 are directly
comparable with the corresponding entries in 
Table B5 of TGB98, modulo the comments made in the
caption to Fig. 1.
The agreement is excellent save for the
WW95$^{-\infty}$ [$^{16}$O,$^{24}$Mg/$^{56}$Fe] entries, for which TGB98 claim
values of -0.21.

\subsection{Elemental vs Isotopic Abundances}
A common theme in the T97 and TGB98 analyses 
is the inherent assumption that isotopic abundance ratios are
equivalent to total elemental abundance ratios - e.g. both assume that
[$^{24}$Mg/$^{56}$Fe]$\equiv$[Mg/Fe], and both 
only consider the individual dominant
isotopes $^{16}$O, $^{24}$Mg, and $^{56}$Fe in their models.  
We wish to remind the reader that this is not entirely correct.

Both groups tie their models to the derived [O,Mg/Fe] Galactic halo
stellar abundances (Figs 15,16 of TGB98; Fig 1 of T97), which reflect the \it
total elemental \rm abundances, whereas their models only consider the dominant
\it isotopes \rm (c.f. Shetrone 1996 for an example of isotopic Mg abundance
determination).  


Table 1 shows that neglecting the ``secondary'' isotopes can lead to an
error of $\sim 7\rightarrow 20$\% in the true IMF-weighted [Mg/Fe]; for the T95
yields, the elemental [Mg/Fe] is $\sim 7$\% greater than the isotopic.
This is of particular interest in that TGB98 claim a 7\%
underproduction of Mg in the T95 yields, a conclusion predicated upon the
(mistaken) assumption that secondary isotopes could be ignored.
Assessing the WW95 yields is somewhat more complicated.
It is apparent that TGB98 have underestimated the true 
WW95$^{+0}$ [Mg/Fe] by $\sim 10$\%, but this
does not entirely compensate
for the 29\% Mg underproduction found by their study.  The WW95 Mg
``problem'' still exists, and is perhaps worse than TGB98 initially suspected.
While TGB98 underestimated the WW95$^{+0}$ [Mg/Fe] by $\sim 10$\%, they have
actually overestimated the [Mg/Fe] for sub-solar metallicities by $\sim 20$\%,
since for [Fe/H]$<+0.0$ the isotopic ratios are actually
greater than the total elemental ratios.

\section{ICM Iron: Type Ia or Type II Supernovae?}
We now present our conclusions regarding the
fractionary contribution of SNe Types Ia and II to the iron content of
the ICM.  The adopted formalism is as outlined in Ishimaru \& Arimoto (1997)
and GLM97.

Fig 1 shows the predicted ICM abundance ratio [$\alpha$/Fe], as a function of
the ICM iron fraction originating in Type Ia SNe.  Taking the observed ICM
[Si/Fe]=+0.14 as canon\footnote{Because Si is determined to an accuracy at
least a factor of two better than either O or Mg, we stress the former.}, 
we would conclude that $\sim 39$\% of the ICM iron came
from Type Ia SNe.  Assuming a steeper IMF ($x=1.70$ vs $x=1.35$), would reduce
this fractionary prediction to $\sim 27$\%.

GLM97 overestimated the Type Ia SNe ICM iron fractionary contribution by $\sim
10$\%, when adopting the T95 yields, by (mistakenly) neglecting the
contribution of the stellar envelope abundances to the total stellar yield.
This $\sim 10$\% effect is demonstrated by the [O/Fe];x=1.35 and [O/Fe]$^{\rm
core}$;x=1.35 curves of Fig 1, and works in the direction of strengthening 
the argument for a mild predominance of Type II SNe-originating iron in the
ICM.


\begin{figure}[!htb]
\plotfiddle{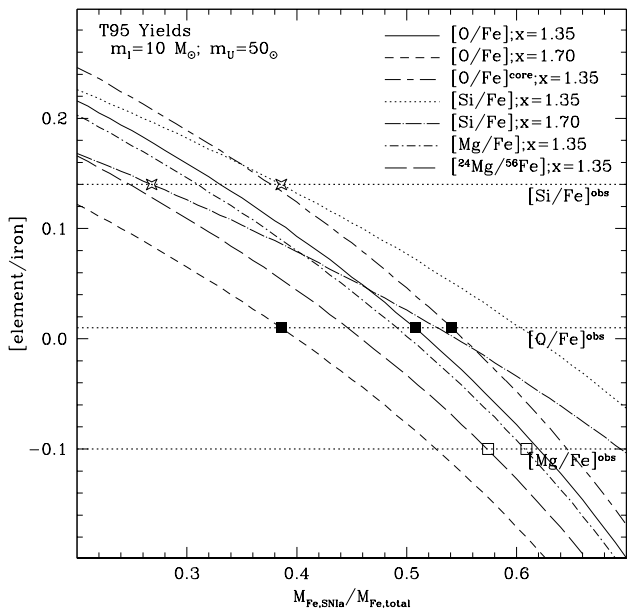}{75truemm}{0}{43}{43}{-150}{-80}
\caption{Ratio of relative predicted ICM abundance (i.e., [element/iron])
as a function of the ICM Type
Ia SNe-originating fraction of Fe.  The T95 Type II SNe yields were employed,
in conjunction with the Nomoto et al. (1997) W7 Type Ia yields.  A \it faux \rm
11 M$_\odot$ entry was generated for the T95 grid, to allow for ease of
comparison with the WW95 grid, by extrapolating
logarithmically from the grid minimum (i.e., 13 M$_\odot$);  elsewhere, though,
linear interpolation between the tabulated entries was adopted.
A Salpeter IMF of slope $x=1.35$, over the mass
range $10\rightarrow 50$ M$_\odot$, was assumed.
The horizontal dotted lines represent
the mean of Mushotzky et al.'s (1996) observed 
\it ASCA \rm SIS abundance data ([Si,O,Mg/Fe] are shown), as
scaled to the meteoritic iron abundance by Ishimaru \& Arimoto (1997).  Where a
model [Si,O,Mg/Fe] curve intersects its corresponding observed ratio, a symbol
is placed to guide the eye.  A range of parameters are covered by the displayed
models, including differing $\alpha$-elements (Si,O,Mg), IMF slope
($x=1.35,1.70$), neglecting (mistakenly) the stellar envelope abundance 
contribution (superscript ``core''), and considering (mistakenly)
only the isotopic ratio ($^{24}$Mg/$^{56}$Fe). See text for further details.
}
\end{figure}

Fig 2 parallels that of Fig 1, but presents predictions for
the full metallicity-dependent grid of WW95.  Again, the [Si/Fe] predictions
imply a \it maximum \rm Type Ia ICM Fe contribution of $\sim 40$\%.

GLM97 (mistakenly) assumed that the extrema in the WW95 IMF-weighted
[$\alpha$/Fe] were given by the [Fe/H]=+0.0 and -4.0 grids - Table 1 and Fig 2
demonstrate that this assumption was incorrect.  On the other hand,
since (i) $\sim$L$_\ast$ ellipticals are the dominant source of ICM metals 
(Gibson \& Matteucci 1997); (ii) the dispersion in ellipticals' stellar [Fe/H] 
distribution is small (Greggio 1997); and, (iii)
peaked about [Fe/H]$\approx -0.3$ (Arimoto et al. 1997), GLM97
would have been better served by considering the WW95 [Fe/H]=-1.0
and +0.0 grids as brackets for the population ultimately responsible for
enriching the ICM.  Doing so would have led to the conclusion that Type Ia SNe
contribute $\sim 5\rightarrow 40$\% of the ICM Fe (middle panel of Fig 2), with
$\sim 20\rightarrow 30$\% being the most probable fraction.


\begin{figure}[!htb]
\plotfiddle{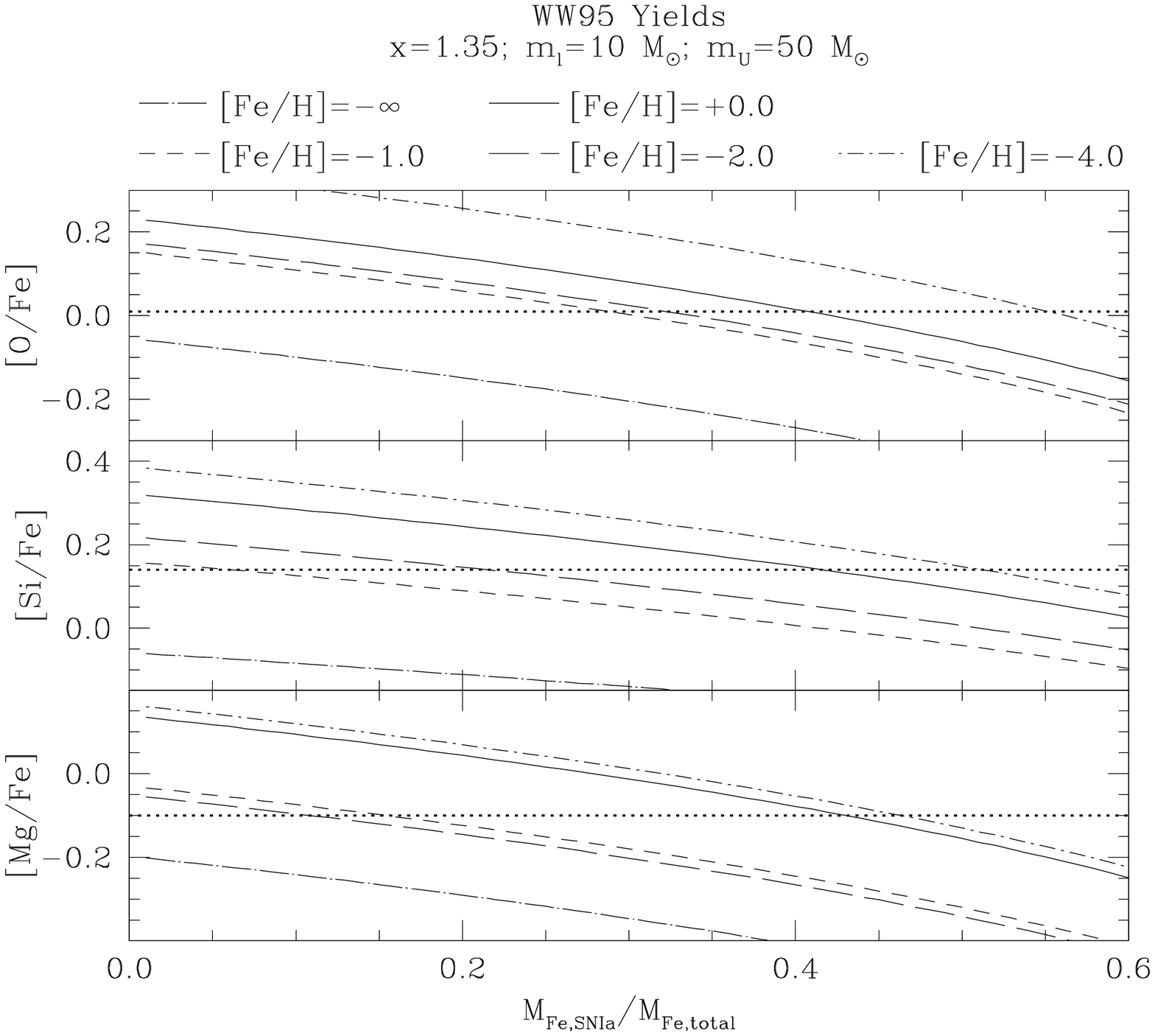}{75truemm}{0}{45}{45}{-150}{-80}
\caption{Ratio of relative predicted ICM abundance (i.e., [element/iron])
as a function of the ICM Type Ia SNe-originating fraction of Fe.  The
metallicity-dependent yields of WW95 were employed; for [Fe/H]$\le -1.0$,
extrapolation below 12 M$_\odot$ was handled as described in the caption to
Figure 1.  Logarithmic extrapolation was used for $m>40$ M$_\odot$, in order to
ensure non-negative yields for all elements, but this does not impact upon our
results.  See text for further details.
}
\end{figure}

\section{Conclusion}
By restricting ourselves to the most accurately determined ICM
$\alpha$-element abundance (i.e., silicon) and adopting
the solar metallicity yields of T95, we conclude that Type Ia SNe
are responsible for $\sim 39/27$\% (for IMF slopes
above $m=10$ M$_\odot$ of $x=1.35/1.70$) of the total ICM
iron budget - i.e., a mild Type II SNe predominance is favored.  Using the
WW95 metallicity-dependent yields only strengthens this conclusion, leading to
a predicted Type Ia fractionary ICM iron contribution of $\sim 0\rightarrow
40$\%, with values in the vicinity of $\sim 20\rightarrow 30$\% the
most likely.  An upper limit to the Type Ia fractionary contribution of $\sim
40$\%, while still allowing a reasonable degree of leeway for the models, does
permit one to exclude several models which predict extreme values (i.e., in
excess of $\sim 80$\%), such as those of Matteucci \& Vettolani (1988) and
Chiosi et al. (1998).



\end{document}